\documentclass[letterpaper,english,aps]{revtex4-1}
\usepackage{lmodern}
\usepackage[T2A,T1]{fontenc}
\usepackage[latin9]{inputenc}
\usepackage{color}
\usepackage{babel}
\usepackage{array}
\usepackage{textcomp}
\usepackage{amsmath}
\usepackage{amssymb}
\usepackage{graphicx}
\usepackage{subscript}
\usepackage[unicode=true,pdfusetitle,
 bookmarks=true,bookmarksnumbered=false,bookmarksopen=false,
 breaklinks=false,pdfborder={0 0 1},backref=false,colorlinks=true]
 {hyperref}
\hypersetup{
 linkcolor=blue, citecolor=blue, urlcolor=blue, pdfstartview=XYZ, plainpages=false, pdfpagelabels}
\usepackage{breakurl}

\makeatletter

\AtBeginDocument{\DeclareFontEncoding{T2A}{}{}}

\newcommand{\lyxmathsym}[1]{\ifmmode\begingroup\def\b@ld{bold}
  \text{\ifx\math@version\b@ld\bfseries\fi#1}\endgroup\else#1\fi}

\providecommand{\tabularnewline}{\\}

\@ifundefined{textcolor}{}
{%
 \definecolor{BLACK}{gray}{0}
 \definecolor{WHITE}{gray}{1}
 \definecolor{RED}{rgb}{1,0,0}
 \definecolor{GREEN}{rgb}{0,1,0}
 \definecolor{BLUE}{rgb}{0,0,1}
 \definecolor{CYAN}{cmyk}{1,0,0,0}
 \definecolor{MAGENTA}{cmyk}{0,1,0,0}
 \definecolor{YELLOW}{cmyk}{0,0,1,0}
}

\@ifundefined{date}{}{\date{}}
\makeatother

\providecommand{\tabularnewline}{\\}

\makeatother

\begin{document}

\title{Spin flip of multiqubit states in discrete phase space}

\author{K.Srinivasan }

\author{G.Raghavan}

\email{gr@igcar.gov.in}

\affiliation{Theoretical studies section, Material Science Group, Indira Gandhi
centre for atomic research, Kalpakkam, Tamilnadu, 603102, India.}

\date{\today}
\begin{abstract}
Time reversal and spin flip are discrete symmetry operations of substantial
import to quantum information and quantum computation. Spin flip arises
in the context of separability, quantification of entanglement and
the construction of Universal NOT gates. The present work investigates
the relationship between the quantum state of a multiqubit system
represented by the Discrete Wigner Function (DWFs) and its spin-flipped
counterpart. The two are shown to be related through a Hadamard matrix
that is independent of the choice of the quantum net used for the
tomographic reconstruction of the DWF. These results would be of interest
to cases involving the direct tomographic reconstruction of the DWF
from experimental data and in the analysis of entanglement related
properties purely in terms of the Discrete Wigner function.
\begin{description}
\item [{PACS~numbers}] 03.65.Ta, 03.67.Mn, 42.79.Ta.{\small \par}
\end{description}
\end{abstract}
\maketitle

\section{Introduction}

Spin flip and time reversal are involution symmetry operations which
frequently arise in quantum information studies. These are anti-unitary
operations which are physically unrealizable in an idealized sense
\citep{PhysRevLett.83.432}. However such operations are critically
important for entanglement detection, quantum optics, quantum computation
and in the definition of certain entanglement measures \citep{Horodecki19961,PhysRevLett.77.1413,PhysRevLett.84.2726,PhysRevLett.80.2245,PhysRevA.67.032307,PhysRevA.68.022318,1.1723701}.
Phase space representation of quantum states through Wigner functions
provides a natural setting for understanding time reversal. However,
in the case of qubit systems the relationship between the Discrete
Wigner function and its spin-flipped/time reversed counterpart is
ill understood, and the present work is an attempt at filling this
gap. To appreciate how spin flip arises in different contexts, we
shall consider the examples cited earlier in some detail. It is generally
a difficult problem to establish whether a mixed state is entangled
or separable. However, the celebrated Peres-Horodecki criterion, based
on positivity of partial transposition (PPT) \citep{Horodecki19961,PhysRevLett.77.1413},
provides the necessary and sufficient condition for the separability
of bipartite systems of dimension $2\otimes2$, $2\otimes3$ and Gaussian
states \citep{PhysRevLett.84.2726}. Partial transposition for a $2\otimes2$
system is defined by $\rho_{ij,kl}^{T}=\rho_{il,kj}$, where $\rho$
is the density matrix of the system. Partial transposition essentially
amounts to a spin flip/reflection of the second particle through a
plane in the Poincaré sphere \citep{PhysRevLett.83.432}. On a single
qubit, this operation is represented by the operator $\sigma_{y}\mathcal{C}$,
where $\mathcal{C}$ is the complex conjugation of the state in the
computational basis and $\sigma_{y}$, the usual Pauli operator. In
the language of quantum maps, $\rho$ is separable if and only if
selective spin flip on the subsystem is a positive map. For a single
qubit system represented by the pure state vector $\left|\psi\right\rangle $,
the spin flipped state is defined by $|\tilde{\psi}>=(-i\sigma_{y})|\psi^{*}>$
and likewise, for the density matrix, it is defined as $\tilde{\rho}=\sigma_{y}\rho^{*}\sigma_{y}$.
Unitary evolution constitutes a completely positive map which takes
quantum states to quantum states and in hindsight, it is not surprising
that only anti-unitary operations are effective for entanglement detection.
In the context of quantum optics, the reflection of an elliptically
polarized photon at a mirror, resulting in a polarization state orthogonal
to it, is analogous to a spin-flip. This operation transforms the
polarization state of the photon to the anti-podal one in the Poincaré
sphere. In quantum computation and quantum simulation, the construction
of Universal NOT (U-NOT) gates for selectively flipping a single qubit,
though not perfectly feasible, is essential. Finally, since entanglement
is viewed as a resource, there is a strong requirement to quantify
it. Several entanglement measures such as negativity, concurrence,
tangles and their generalizations to qudit and multiqubit systems
have been proposed in the literature \citep{PhysRevLett.80.2245,PhysRevA.58.883,PhysRevA.65.032314,1497700,PhysRevA.63.044301,PhysRevA.61.052306}.
The definition of these measures critically hinge on the spin flip
operation. For example, for arbitrary bipartite mixed states, Wootters
\citep{PhysRevLett.80.2245} has derived an entangled measure called
concurrence, given by the expression:

\begin{equation}
C(\rho)=max\left\{ 0,\sqrt{\lambda}_{1}-\sqrt{\lambda}_{2}-\sqrt{\lambda}_{3}-\sqrt{\lambda}_{4}\right\} \label{eq:Concurrence}
\end{equation}

where the $\lambda's$ are the eigenvalues of $R=\rho\tilde{\rho}$
and $\tilde{\rho}$ is the spin flipped state defined as:
\begin{equation}
\tilde{\rho}=\left(\sigma_{y}\otimes\sigma_{y}\right)\rho^{*}\left(\sigma_{y}\otimes\sigma_{y}\right)\label{eq:rho-tilde}
\end{equation}

Concurrence is an entanglement monotone viz., a quantity which is
invariant under local quantum operations and classical communication.
Given the usefulness of this quantity for entanglement quantification,
there have been attempts at generalizing its definition for systems
of higher dimensions and for multipartite states \citep{PhysRevA.67.032307,PhysRevA.64.042315}.
For instance, the n-qubit concurrence $C_{n}($$\rho)$ is defined
in terms of the eigenvalues of the $R=\rho\tilde{\rho}$, where n-qubit
spin flip operation is defined as:
\begin{equation}
\tilde{\rho}=\sigma_{y}^{\otimes n}\rho^{*}\sigma_{y}^{\otimes n}\label{eq:rhotild}
\end{equation}

Though attempts at investigating entanglement have largely been based
on representing the state through the density matrix, it by no means
the unique way of doing so. In fact, alternate representations of
the state through quasi-probability distributions such as the Wigner
functions and through Stokes vectors are prevalent in quantum optics.
Both these quantities can be readily reconstructed from measured data.
These quantities are therefore valid representations of the state
in their own right. Wigner functions have been used to investigate
quantum states of light such as entangled, squeezed and photon-added
coherent states \citep{eltit,Zavatta22102004}. These are normalized
and real valued functions which can however assume negative values
over restricted regions of the domain, and are therefore called quasi-probability
distributions \citep{PhysRev.40.749,Hillery1984121}. Indeed, the
negativity of the Wigner function is taken to be a signature of the
non-classicality of the state and is indicative of quantum interference
effects \citep{1464-4266-6-10-003}. Besides providing a visual presentation
of these effects, Wigner functions have been used for investigating
quantum dynamics, quantum random walks, decoherence and entanglement
detection of continuous systems \citep{PhysRevLett.84.2726,RevModPhys.75.715}.
The representation of the state through Wigner functions has the advantage
that time reversal has a very transparent interpretation. The extension
of the Peres-Horodecki criterion to bipartite Gaussian states by Simon
\citep{PhysRevLett.84.2726} is based on the critical observation
that transposition in the continuous case, is geometrically interpreted
as a mirror reflection in the phase space. This is evident from the
observation that transformation $\rho\rightarrow\rho^{T}$, corresponds
to the associated Wigner function transforming as: $W(q,p)\rightarrow W(q,-p)$.
Since quantum information and computation applications generally use
qubits, Discrete Wigner functions (DWFs) have been investigated in
the literature. Unlike continuous Wigner functions, Discrete Wigner
functions are not unique, and since the underlying field is discrete,
certain restrictions are imposed. Constructions based on discrete
$2d\times2d$ \citep{Hannay1980267,Bianucci2002353} and $d\times d$
grids \citep{Cohen1986,Feynman1987-FEYNP,PhysRevB.10.3700,Galetti1992513,0305-4470-21-13-012}
have been proposed in the literature. The DWF construction for the
odd dimensional systems are given by Gross \citep{gross1.2393152}
and Chaturvedi et al \citep{ref1}. Unlike the continuous case, there
is no clear link between negativity and non-classicality in the case
of DWFs. Even for maximally entangled bipartite states, the DWF is
not necessarily negative for every choice of the quantum net. Despite
this limitation, DWFs have been used to investigate stabilizer states
which arise in the context of error correcting codes \citep{PhysRevA.73.012301,PhysRevA.72.012309},
quantum optics \citep{PhysRevA.41.5156}, quantum state tomography
\citep{PhysRevA.64.012106,PhysRevA.53.2998} and teleportation \citep{PhysRevA.65.062311}.
The construction proposed by Wootters \citep{Wootters1986,Wootters:2004:PQP:1014615.1014629,WOOTTERS19871},
Gibbons et al., \citep{PhysRevA.70.062101} for $d\times d$ is particularly
well suited for these studies and we shall use the same in the present
work. In investigations involving the representation in terms of the
DWF $W$, the spin flipped DWF $\tilde{W}$ is required to qualitatively
and quantitatively evaluate the state. Given this background, there
is strong motivation to examine the relationship between $W$ and
$\tilde{W}$ and to provide a prescription for computation of the
latter from the former. Towards this end, in the present work, we
exhibit an elegant relationship between $W$ and $\tilde{W}$. We
show that the $n$ qubit DWF is related to its spin flipped counterpart
through an $N^{2}\times N^{2}$ Hadamard matrix, where $N=2^{n}$.
Since the construction of the DWF depends on the Mutually Unbiased
Basis sets (MUBS) that are employed, the DWF that is associated with
a given density matrix is no longer unique as in the continuous case.
Different bases sets lead to different choices of the so-called quantum
net. It is therefore essential to show that the construction is independent
of the quantum net and the proof for a general $n$ - qubit system
is also provided. The rest of the paper is structured as follows:
Section II provides a quick overview of the Discrete Wigner function
and Wootters' construction. The spin flipped DWF is derived in two
steps in Section III. In section A, we show that $W^{(*)}$, the DWF
of $\rho^{*}$ is related to $W$ through a Hadamard matrix and III-B
and we show that this is independent of the choice quantum net. In
III-C the DWF of the spin flipped density matrix is obtained by effecting
a generalized shift in $W^{(*)}$. The final transformation matrix
relating $\tilde{W}$ to $W$ also turns out to be Hadamard matrix
which, is once again independent of the quantum net. Section IV-A
illustrates the method for a single qubit system. Section IV-B outline
the derivation for the two qubit case. The final section provides
the conclusions with some brief remarks.

\section{Discrete Wigner function}

This construction applies to Hilbert space of dimension $N$ where,
$N$ is prime number or $N=r^{n}$, where $r$ is prime. The DWF is
then defined over an $N\times N$ array of discrete points defined
over a Galois field $\mathcal{F_{N}}$. The points are labelled by
an ordered set $\alpha=(p,q)$. Since our interest is primarily in
qubit states, the dimension of the Hilbert's space is $N=2^{n}.$
This construction uses the notion of lines, translational covariance
and marginals in analogy with Wigner functions of continuous systems.
A line is defined by as collection of points solving the equation
$aq+bp=c$. A set of lines with identical $a$ and $b$ but with a
different $c$ are said to be parallel to each other. In Euclidean
geometry, parallel lines do not share any point and non-parallel lines
intersect at exactly one point. To ensure this, the underlying finite
field structure has to be correctly chosen \citep{PhysRevA.70.062101}.
In this construction, there are $N+1$ sets of parallel lines with
exactly $N$ lines in each of them. Each set of parallel lines is
called a striation. Crucially, a rank one projector $P_{\lambda}=\left|\lambda\right\rangle \left\langle \lambda\right|$
is associated with each line $\lambda$, and the sum of the the Wigner
function over all points $\alpha$ contained in the line is required
to satisfy the condition:

\begin{equation}
\underset{\alpha\in\lambda}{\sum}W_{\alpha}=Tr\left(\rho P_{\lambda}\right)
\end{equation}

where $\rho$ is quantum state. Self adjoint operators $\hat{A}_{\alpha}$
with unit trace, having the property: $Tr(\hat{A}_{\alpha})=1$ and
$Tr(\hat{A}_{\alpha}\hat{A}_{\beta})=N\delta_{\alpha\beta}$, are
associated with each point of the lattice. With this association,
the density operator may be written as:

\begin{equation}
\hat{\rho}=\underset{\alpha}{\sum}W_{\alpha}\hat{A}_{\alpha}\label{eq:rhoW}
\end{equation}

Thus, the $\hat{A}_{\alpha}$ operators constitute an orthonormal
basis and Wigner function are expansion coefficients. The projection
operators are related to the operators $\hat{A}_{\alpha}$ as:

\begin{equation}
P_{\lambda}=\frac{1}{N}\underset{\alpha\in\lambda}{\sum}A_{\alpha}
\end{equation}

This set of orthonormal vectors $\{\left|\lambda\right\rangle _{j}^{\kappa}\}$
associated with the striation $\kappa$ constitutes a basis set. There
are therefore $N+1$ basis sets $B^{\kappa}$: $\{\lambda_{1}^{\kappa},\lambda_{2}^{\kappa},.....\lambda_{N}^{\kappa}\}$
with the property:

\begin{equation}
\left|\left\langle \lambda_{j}^{k}\right|\left.\lambda_{j^{'}}^{\kappa^{'}}\right\rangle \right|^{2}=\frac{1}{N}\left(1-\delta_{\kappa,\kappa^{'}}\right)+\delta_{\kappa,\kappa^{'}}\delta_{j,j^{'}}
\end{equation}

Such a set therefore constitutes a mutually unbiased basis set (MUBS)
\citep{0305-4470-14-12-019}\citep{Wootters1986}. The expression
for the Wigner function $W_{\alpha}$ at every point $\alpha$ can
now be readily supplied:

\begin{equation}
W_{\alpha}=\frac{1}{N}Tr\left(\hat{\rho}\hat{A_{\alpha}}\right)\label{eq:WAalpha}
\end{equation}
where: 

\begin{equation}
\hat{A_{\alpha}}=\frac{1}{N}\left(\underset{\lambda_{j}^{\kappa}\ni\alpha}{\sum}\hat{P}_{j}^{\kappa}-I\right)\label{eq:PPoperator}
\end{equation}

The important point is that the outcome probabilities associated with
projective measurements along a line are:

\begin{equation}
p_{j}^{\kappa}=Tr\left(\hat{\rho}\hat{P}_{j}^{\kappa}\right)
\end{equation}
and the Wigner function itself can be tomographically constructed
through repeated measurements along all the lines as:

\begin{equation}
W_{\alpha}=\frac{1}{N}\left(\underset{\lambda_{j}^{\kappa}\ni\alpha}{\sum}p_{j}^{\kappa}-I\right)
\end{equation}

To ensure translational covariance of the states, we associate a unitary
operator $\hat{T}_{(x,y)}$ with every phase space point with the
condition that the composition law $\hat{T}_{\alpha}\hat{T}_{\beta}=T_{\alpha+\beta}$
applies. The general unitary translation operator in phase space can
written as,
\begin{equation}
T_{(q,p)}=\sigma_{x}^{q_{1}}\sigma_{z}^{p_{1}}\otimes\sigma_{x}^{q_{2}}\sigma_{z}^{p_{2}}...\otimes\sigma_{x}^{q_{n}}\sigma_{z}^{p_{n}}\label{eq:translation}
\end{equation}

As in the continuous case, the DWF is translationally covariant. Let
$W$ and $W'$ be the DWF associated with the states $\rho$ and $\rho'$.
If $\rho$ and $\rho'$ are related by the unitary translation operator
$T_{\beta}$ by,

\begin{equation}
\rho'=\hat{T}_{\beta}\rho\hat{T}_{\beta}^{\dagger}
\end{equation}

then : 
\begin{equation}
W'_{\alpha}=W_{\alpha+\beta}
\end{equation}

Thus, $W'$ is obtained by shifting each element of $W$ by an amount
$\beta$

\section{Spin flipped DWF of a multiqubit system}

To derive a relationship between the DWF $W$ and its spin-flipped
counterpart $\tilde{W}$, we begin by observing that the spin flip
operation on the density matrix is a two step process: the first step
involves the complex conjugation of $\rho$ in the computational basis
and the next one entails the application of the translational operators
$\sigma_{y}^{\otimes n}$ to it. Consequently, the computation of
the spin flipped DWF $\tilde{W}$ can be carried out in two steps.
We denote the DWF of $\rho^{*}$ by $W^{(*)}$ (not to confused with
complex conjugation of $W$, which in any case is real valued). In
the first step, we derive an expression for $W^{\left(*\right)}$
in terms of $W$. A shift associated with the translation $\sigma_{y}^{\otimes n}$
of $\rho^{*}$ is then effected by subjecting $W^{\left(*\right)}$
to a corresponding shift, to obtain $\tilde{W}$.

\subsection{Derivation of $W^{(*)}$ for the multiqubit state}

If the system is represented by a state vector or a density matrix,
then complex conjugation is straight forward. The procedure is however
not obvious when the system is represented by the DWF. To elucidate
this, we exploit the relationship between $W$ and $\rho$ and write
$W_{\alpha}^{(*)}$ as:

\begin{equation}
W_{\alpha}^{(*)}=\frac{1}{N}Tr\left(\rho^{*}A_{\alpha}\right)\label{eq:wstar1}
\end{equation}

since $A_{\alpha}$ are self adjoint operators. Now taking the complex
conjugate of Eq (\ref{eq:rhoW}) we have:
\begin{equation}
\rho^{*}=\underset{\beta}{\sum}W_{\beta}A_{\beta}^{*}\label{eq:rhostar}
\end{equation}

substituting Eq.(\ref{eq:wstar1}) in Eq. (\ref{eq:rhostar}), thus:
\begin{equation}
W_{\alpha}^{(*)}=\frac{1}{N}Tr\left[\left(\underset{\beta}{\sum}W_{\beta}A_{\beta}^{*}\right)A_{\alpha}\right]\label{eq:wstar2}
\end{equation}

taking the trace operation inside the summation:
\begin{equation}
W_{\alpha}^{(*)}=\frac{1}{N}\underset{\beta}{\sum}W_{\beta}Tr\left(A_{\alpha}A_{\beta}^{*}\right)\label{eq:wstar4}
\end{equation}

writing the Wigner elements as a column vector and the Wigner elements
in terms of the field labels:
\begin{equation}
\left(\begin{array}{c}
W_{0,0}^{(*)}\\
\vdots\\
W_{1,0}^{(*)}\\
\vdots\\
W_{\bar{\omega},\bar{\omega}}^{(*)}
\end{array}\right)=\frac{1}{N}\left(\begin{array}{ccccc}
Tr\left(A_{0,0}A_{0,0}^{*}\right) & Tr\left(A_{0,0}A_{0,1}^{*}\right) & \cdots & \cdots & Tr\left(A_{0,0}A_{\bar{\omega},\bar{\omega}}^{*}\right)\\
Tr\left(A_{0,1}A_{0,0}^{*}\right) & Tr\left(A_{0,1}A_{0,1}^{*}\right) & \cdots & \cdots & Tr\left(A_{0,1}A_{\bar{\omega},\bar{\omega}}^{*}\right)\\
\vdots & \vdots & \vdots & \vdots & \vdots\\
\vdots & \vdots & \vdots & \vdots & \vdots\\
Tr\left(A_{\bar{\omega},\bar{\omega}}A_{0,0}^{*}\right) & Tr\left(A_{\bar{\omega},\bar{\omega}}A_{0,1}^{*}\right) & \cdots & \cdots & Tr\left(A_{\bar{\omega},\bar{\omega}}A_{\bar{\omega},\bar{\omega}}^{*}\right)
\end{array}\right)\left(\begin{array}{c}
W_{0,0}\\
\vdots\\
W_{1,0}\\
\vdots\\
W_{\bar{\omega},\bar{\omega}}
\end{array}\right)\label{eq:wstarmat}
\end{equation}

This equation can be written in a compact form:
\begin{equation}
W^{(*)}=SW\label{eq:wstarf}
\end{equation}

where $S$ is a $N^{2}\times N^{2}$ matrix. Thus, Eq. \ref{eq:wstarf}
helps us to implement the complex conjugation operation for the DWFs.
In the next section we show that the matrix $S$ is a Hadamard matrix
and that it is independent of the quantum net.

\subsection{Proof that S is a Hadamard Matrix and that it is independent of
the quantum net}

While the DWF depends on the specific choice of the quantum net, we
show that the transformation matrix $S$ is itself independent of
this choice. Thus, we shall show that a single Hadamard matrix $S$
is sufficient for transforming $W$, obtained using any quantum net,
to the corresponding $W^{(*)}$. It may easily be verified that the
transformation matrix $S$ is made of $N\times N$ blocks, of size
$N\times N$. We shall label the blocks of $S$ as $S_{ij}$ where
the suffixes are the block indices. Note however that $S(k,l)$ denote
the matrix elements. For example, the first block $S_{11}$ is a
$N\times N$ matrix formed by varying the elements $\alpha$ and $\beta$
in $Tr(A_{\alpha}A_{\beta}^{*})$ over the points in the first line
of the first striation. We now start by writing the explicit equation
for $Tr(A_{\alpha}A_{\beta}^{*})$. Using Eq(9), 
\begin{align}
Tr(A_{\alpha}A_{\beta}^{*})= & Tr\left[\left(\underset{\lambda\ni\alpha}{\sum}Q(\lambda)-I\right)\left(\underset{\lambda'\ni\beta}{\sum}Q^{*}(\lambda')-I\right)\right]\nonumber \\
= & Tr\left\{ \left[Q(\lambda_{1}^{x_{1}})+\dots+Q(\lambda_{N+1}^{x_{N+1}})\right]\left[Q^{*}(\lambda_{1}^{y_{1}})+\dots+Q^{*}(\lambda_{N+1}^{y_{N+1}}\right]\right\} -N-2\nonumber \\
= & T_{\alpha\beta}-N-2\label{eq:T_alpha}
\end{align}

where the term $T_{\alpha\beta}$ contains $(N+1)^{2}$ terms of the
form $Tr[Q(\lambda_{i}^{x_{i}})Q^{*}(\lambda_{j}^{y_{j}})]$. The
$Q(\lambda_{i}^{x_{i}})s$ are rank one projectors associated with
the $x_{i}$$^{th}$ line in the $i$$^{th}$ striation. In the $N+1$
MUBS, the first basis sets can always be taken to have real elements
and hence complex conjugation does not alter them. Therefore, for
these two cases, $Tr[Q(\lambda_{i}^{x_{i}})Q^{*}(\lambda_{i}^{x_{i}})]=1$
(where $i\in[1,2]$) as $Q^{*}(\lambda_{i}^{x_{i}})=Q(\lambda_{i}^{x_{i}})$.
The other $N-1$ bases have complex entries and are closed under complex
conjugation i.e,. complex conjugation takes each basis vector to another
one one orthogonal to it. Therefore, $Tr[Q(\lambda_{i}^{x_{i}})Q^{*}(\lambda_{i}^{x_{i}})]=0$
for $i\in[3,4,\dots N+1]$. If however, $Q(\lambda_{i}^{x_{i}})$
and $Q^{*}(\lambda_{j}^{y_{j}})$ belong to different striations i.e,.
different bases sets, then, $Tr[Q(\lambda_{i}^{x_{i}})Q^{*}(\lambda_{j}^{y_{j}})]=\frac{1}{N}$
($i\ne j$). It is important to note that for all these cases, $Tr[Q(\lambda_{i}^{x_{i}})Q^{*}(\lambda_{j}^{y_{j}})]$
is independent of the quantum net. It is essentially arises from the
fact that DWF is translationally covarient. 

Consider the case $\alpha=\beta$ in $Tr(A_{\alpha}A_{\beta}^{*})$,
which are the set of all diagonal elements of the matrix $S$. From
Eq (\ref{eq:T_alpha}) $Tr(A_{\alpha}A_{\beta}^{*})=T_{\alpha\beta}-N-2$
, where the term $T_{\alpha\beta}$ has $(N+1)^{2}$ trace terms.
There are $N(N+1)$ trace terms $Tr[Q(\lambda_{i}^{x_{i}})Q^{*}(\lambda_{j}^{y_{j}})]=\frac{1}{N}$
for which $i\neq j$, that is $Q(\lambda_{i}^{x_{i}})$ and$Q^{*}(\lambda_{j}^{y_{j}})$
are from different striations, so their value become $N(N+1)\frac{1}{N}=N+1$.
In the remaining $N+1$ terms, the value of $Tr[Q(\lambda_{i}^{x_{i}})Q^{*}(\lambda_{i}^{x_{i}})]$
is $1$ for the first two striations and $0$ for the other $N-1$
striations. Therefore $Tr(A_{\alpha}A_{\alpha}^{*})=1$ for all $\alpha$,
so that all the diagonal entries of the $S$ matrix are all equal
to $1$.

Next, consider the case in which $\alpha$ and $\beta$ belong to
the same line in the first striation. This spans all the diagonal
blocks of the $S$ matrix. The variable $T_{\alpha\beta}$ in $Tr(A_{\alpha}A_{\beta}^{*})$
is $N(N+1)$ times $\frac{1}{N}$, since $Tr[Q(\lambda_{i}^{x_{i}})Q^{*}(\lambda_{j}^{y_{j}})]=\frac{1}{N}$,
for which, $i\neq j$. Since $\alpha$ and $\beta$ are from the same
striation, for each value of $\alpha$ and $\beta$ , there are two
trace terms which contribute $1$ and other terms are zero. So in
this case $Tr(A_{\alpha}A_{\beta}^{*})=1$. Therefore, all the diagonal
blocks of $S$ are $N\times N$ matrices with unit entries.

As the last case, consider $\alpha$ and $\beta$ belonging to the
different lines of the first striation. This condition spans all the
off-diagonal blocks of the $S$ matrix. Here too, the calculation
of $T_{\alpha\beta}$ gives $N(N+1)$ times $\frac{1}{N}$ for the
$i\neq j$ case. Since $\alpha$ and $\beta$ are from two distinct
lines of the first striation $Tr[Q(\lambda_{1}^{x_{1}})Q^{*}(\lambda_{1}^{y{}_{1}})]=0$
,$\forall\alpha,\beta$. For the given $\alpha$, if $\beta$ runs
over all the points in the given striation, it gives a particular
row in that block. In that row $\frac{N}{2}$ entries are $+1$s and
$\frac{N}{2}$ are $-1$s. 

Thus, in $S$, all the diagonal blocks are $N\times N$ matrices have
entries which are $+1$ and all the off-diagonal blocks are the $N\times N$
matrix with each row containing an equal number of $+1s$ and $-1s$.
Further, the rows and columns of $S$ are orthogonal to each other.We
thus see that $S$ is a Hadamard matrix. We not that these observations
are not specific to any quantum net, implying that $S$ is independent
of the same.

\subsection{The spin flipped Wigner function $\tilde{W}$}

The next step now is to obtain the spin flipped DWF $\tilde{W}$ from
$W^{(*)}$. In terms of density matrices, the spin flip operation
is defined as $\tilde{\rho}=\sigma_{y}^{\otimes n}\rho^{*}\sigma_{y}^{\otimes n}$.
To find the $\tilde{W}$, the Pauli operators acting on the individual
qubits are effectively translation operators $T_{\beta}$ acting on
$W^{(*)}$, that shift each element of $W^{(*)}$ in the phase space
by an amount $\beta$. We can realize this transformation by a $N^{2}\times N^{2}$
matrix $T$ acting on $W^{(*)}$ (arranged as a column vector) to
obtain $\tilde{W}$:
\begin{equation}
\tilde{W}=TW^{(*)}
\end{equation}

$T$ depends on the basis choice of the underlying Galois field. Using
the Eq (\ref{eq:wstarf}) we can write $\tilde{W}$ directly in terms
of $W$ as:
\begin{equation}
\tilde{W}=TSW
\end{equation}
 $T$ acting on $S$ merely interchanges the rows, and so the resultant
matrix is again a Hadamard matrix $H=TS$. Thus, the spin flipped
state takes the form:
\begin{equation}
\tilde{W}=HW
\end{equation}

This completes the derivation of $\tilde{W}$ from $W$ and the proof
that the Hadamard matrix $H$ is quantum net independent.

\section{Illustration of the spin flip operation for one and two qubit DWFs}

\subsection{Illustration for a one qubit system}

In order to clarify the procedure for computing $\tilde{W}$, let
us consider the spin flip operation on a one qubit system. For this
case, the axes of the discrete phase space are labelled by the Galois
field elements $\mathbb{F}_{2}=\{0,1\}$. In the $2\times2$ phase
space, there are $3$ striations and each striation contains two lines,
having totally $6$ lines. The MUBs associated with the $3$ striations
are,
\begin{eqnarray*}
B_{1}=\left(\begin{array}{cc}
1 & 0\\
0 & 1
\end{array}\right)\;\; & B_{2}=\frac{1}{\sqrt{2}}\left(\begin{array}{cc}
1 & 1\\
1 & -1
\end{array}\right)\;\; & B_{3}=\frac{1}{\sqrt{2}}\left(\begin{array}{cc}
1 & 1\\
-i & i
\end{array}\right)
\end{eqnarray*}

where the matrices are the basis sets and the columns of $B_{i}$
are the mutually orthogonal basis vectors. The first two basis sets
$B_{1}$ and $B_{2}$ are not altered by complex conjugation, but
for the third basis set $B_{3}$, complex conjugation interchanges
columns $1$ and $2$. From Eq (\ref{eq:wstarf}) we have $W^{(*)}=SW$.
$S$ can be therefore be written as,
\begin{equation}
S=\frac{1}{2}\left(\begin{array}{cccc}
Tr(A_{0,0}A_{0,0}^{*})\, & Tr(A_{0,0}A_{0,1}^{*})\, & Tr(A_{0,0}A_{1,0}^{*})\, & Tr(A_{0,0}A_{1,1}^{*})\\
Tr(A_{0,1}A_{0,0}^{*})\, & Tr(A_{0,1}A_{0,1}^{*})\, & Tr(A_{0,1}A_{1,0}^{*})\, & Tr(A_{0,1}A_{1,1}^{*})\\
Tr(A_{1,0}A_{0,0}^{*})\, & Tr(A_{1,0}A_{0,1}^{*})\, & Tr(A_{1,0}A_{1,0}^{*})\, & Tr(A_{1,0}A_{1,1}^{*})\\
Tr(A_{1,1}A_{0,0}^{*})\, & Tr(A_{1,1}A_{0,1}^{*})\, & Tr(A_{1,1}A_{1,0}^{*})\, & Tr(A_{1,1}A_{1,1}^{*})
\end{array}\right)=\frac{1}{2}\left(\begin{array}{cc}
S_{11} & S_{12}\\
S_{21} & S_{22}
\end{array}\right)
\end{equation}

where $S$ is written as a block matrix with $2\times2$ blocks, each
block being a $2\times2$ matrix. By examining Eq (\ref{eq:T_alpha}),
clearly, $Tr(A_{\alpha}A_{\beta}^{*})=T_{\alpha\beta}-4$ and,
\begin{itemize}
\item For points $\alpha=\beta$, i.e for the diagonal entries of $S_{11}$
and $S_{22}$, $Tr[Q(\lambda_{i}^{x_{i}})Q^{*}(\lambda_{i}^{x_{i}})]=1$
for $i=1,2$ and $Tr[Q(\lambda_{i}^{x_{i}})Q^{*}(\lambda_{i}^{y_{i}})]=0$
for $i=3$. Therefore, $T_{\alpha\alpha}=[2(2+1)\frac{1}{2}]+1+1=5$.
So, $Tr(A_{\alpha}A_{\alpha}^{*})=1$. Therefore, for all the diagonal
entries $Tr(A_{\alpha}A_{\alpha}^{*})=1$.
\item If $\alpha$ and $\beta$ belong to the same line of the first striation
then, $Tr[Q(\lambda_{1}^{x_{1}})Q^{*}(\lambda_{1}^{x_{1}})]=1$, whenever
$Tr[Q(\lambda_{2}^{x_{2}})Q^{*}(\lambda_{2}^{x_{2}})]=1$ then $Tr[Q(\lambda_{3}^{x_{3}})Q^{*}(\lambda_{3}^{x_{3}})]=0$
and vice versa. Therefore, $T_{\alpha\beta}=[2(2+1)\frac{1}{2}]+1+1=5$.
So $Tr(A_{\alpha}A_{\beta}^{*})=1$. This argument holds for all the
diagonal blocks. Thus, the diagonal blocks $S_{11}$ and $S_{22}$
are of the form:$\left(\begin{array}{cc}
1 & 1\\
1 & 1
\end{array}\right).$
\item For the off diagonal blocks, $\alpha$ and $\beta$ belong to different
lines of the first striation and $Tr[Q(\lambda_{1}^{x_{1}})Q^{*}(\lambda_{1}^{y_{1}})]=0$
for all $\alpha$ and $\beta$. If points $\alpha$ and $\beta$ belong
to the same line of the third striation, then, $Tr[Q(\lambda_{3}^{x_{3}})Q^{*}(\lambda_{3}^{x_{3}})]=0$
. Therefore, $T_{\alpha\beta}=[2(2+1)\frac{1}{2}]=3$. So $Tr(A_{\alpha}A_{\beta}^{*})=-1$,
otherwise $Tr(A_{\alpha}A_{\beta}^{*})=1$. For example $\alpha=(0,0)$
and $\beta=(1,1)$ belong to different lines of the first striation
and same line of the third striation, so $Tr(A_{\alpha}A_{\beta}^{*})=-1$. 
\end{itemize}
The transformation matrix $S$ thus takes the form,
\begin{equation}
S=\frac{1}{2}\left(\begin{array}{cccc}
1 & 1 & 1 & -1\\
1 & 1 & -1 & 1\\
1 & -1 & 1 & 1\\
-1 & 1 & 1 & 1
\end{array}\right)
\end{equation}

In the construction of this matrix, we have not used any particular
quantum net, hence $S$ is independent of the same. The matrix $S$
has the following properties:
\begin{enumerate}
\item The elements of the $S$ are either $\pm\frac{1}{N}$, where $N=2^{n}$. 
\item Two different rows(columns) are orthogonal to each other.
\item Determinant of the $S$ is $-1$. For a general $n$ qubit system,
the determinant of $S$ is equal to $(-1)^{n}$.
\item It is a self-inverse. 
\end{enumerate}
therefore, the DWF of the complex conjugated state can be written
as,
\begin{equation}
\left(\begin{array}{c}
W_{00}^{(*)}\\
W_{01}^{(*)}\\
W_{10}^{(*)}\\
W_{11}^{(*)}
\end{array}\right)=\frac{1}{2}\left(\begin{array}{cccc}
1 & 1 & 1 & -1\\
1 & 1 & -1 & 1\\
1 & -1 & 1 & 1\\
-1 & 1 & 1 & 1
\end{array}\right)\left(\begin{array}{c}
W_{00}\\
W_{01}\\
W_{10}\\
W11
\end{array}\right)\label{eq:oneWstar1}
\end{equation}

The next task is to perform the unitary translational operation on
$W^{(*)}$, to complete the spin flip operation. In the equation $\tilde{\rho}=\sigma_{y}\rho*\sigma_{y}^{\dagger}$,
the Pauli's operator $\sigma_{y}$ is essentially a translational
operator $T_{\beta}=\sigma_{y}$, which translates every point in
the DWF $W^{(*)}$ by an amount $\beta$. The spin flipped state is
hence given by $\tilde{W_{\alpha}}=W_{\alpha+\beta}^{(*)}$. From
Eq (\ref{eq:translation}), it is seen that the general translational
operator on phase space can be written as $T_{(q,p)}=\sigma_{x}^{q}\sigma_{z}^{p}$.
From this equation, $\sigma_{y}$ may be expressed as $T_{\beta}=\sigma_{y}=-i(\sigma_{x}^{1}\sigma_{z}^{1})=T_{(1,1)}$,
where $-i$ is the phase factor. Therefore, the spin flip operation
is carried out by translating every element of the phase space by
an amount $\beta=(1,1)$. This translation can also be represented
by a matrix :
\begin{equation}
\left(\begin{array}{c}
\tilde{W}_{00}\\
\tilde{W}_{01}\\
\tilde{W}_{10}\\
\tilde{W}_{11}
\end{array}\right)=\left(\begin{array}{cccc}
0 & 0 & 0 & 1\\
0 & 0 & 1 & 0\\
0 & 1 & 0 & 0\\
1 & 0 & 0 & 0
\end{array}\right)\left(\begin{array}{c}
W_{00}^{(*)}\\
W_{01}^{(*)}\\
W_{10}^{(*)}\\
W_{11}^{(*)}
\end{array}\right)\label{eq:oneWt1}
\end{equation}

Therefore $\tilde{W}$ may be written as $\tilde{W}=TW$. We know
that $W^{(*)}=SW$, hence using Eq (\ref{eq:oneWstar1}), and Eq (\ref{eq:oneWt1}),
we may write, 

\begin{equation}
\left(\begin{array}{c}
\tilde{W}_{00}\\
\tilde{W}_{01}\\
\tilde{W}_{10}\\
\tilde{W}_{11}
\end{array}\right)=\frac{1}{2}\left(\begin{array}{cccc}
0 & 0 & 0 & 1\\
0 & 0 & 1 & 0\\
0 & 1 & 0 & 0\\
1 & 0 & 0 & 0
\end{array}\right)\left(\begin{array}{cccc}
1 & 1 & 1 & -1\\
1 & 1 & -1 & 1\\
1 & -1 & 1 & 1\\
-1 & 1 & 1 & 1
\end{array}\right)\left(\begin{array}{c}
W_{00}\\
W_{01}\\
W_{10}\\
W11
\end{array}\right)\label{eq:oneWT2}
\end{equation}

or 
\begin{equation}
\tilde{W}=HW
\end{equation}

The product $H=TS$, is also a Hadamard matrix as explained earlier.

\subsection{Spin flipped DWF of a two qubit system}

To appreciate the general results obtained for the multiqubit state,
it would be helpful to additionally consider the two qubit case in
some detail. The points in the axis of two qubit discrete phase space
are labelled by the finite field $\mathbf{\mathbb{\mathbb{F}}_{4}=\{}0,1\mathbf{,\omega,\bar{\omega}\}}$.
The MUBS associated with this discrete phase space is given by:
\begin{eqnarray*}
B_{1}=\left(\begin{array}{cccc}
1 & 0 & 0 & 0\\
0 & 1 & 0 & 0\\
0 & 0 & 1 & 0\\
0 & 0 & 0 & 1
\end{array}\right) & B_{2}=\frac{1}{2}\left(\begin{array}{cccc}
1 & 1 & 1 & 1\\
1 & -1 & 1 & -1\\
1 & 1 & -1 & -1\\
1 & -1 & -1 & 1
\end{array}\right) & B_{3}=\frac{1}{2}\left(\begin{array}{cccc}
1 & 1 & 1 & 1\\
-i & i & -i & i\\
i & i & -i & -i\\
1 & -1 & -1 & 1
\end{array}\right)\\
B_{4}=\frac{1}{2}\left(\begin{array}{cccc}
1 & 1 & 1 & 1\\
1 & -1 & 1 & -1\\
i & i & -i & -i\\
-i & i & i & -i
\end{array}\right) & B_{5}=\frac{1}{2}\left(\begin{array}{cccc}
1 & 1 & 1 & 1\\
-i & i & -i & i\\
1 & 1 & -1 & -1\\
i & -i & -i & i
\end{array}\right)
\end{eqnarray*}

As in the single qubit case, each matrix is a basis set and the columns
are the basis vectors. Since the first two basis sets do not have
any complex entries, complex conjugation does not alter them. But
the last three bases have complex entries and complex conjugation
takes a vector into some other vector in the same basis. For two qubit
systems, the discrete phase space is a $4\times4$ array of points
having 16 entries, Writing $W$ as a $16\times1$ column vector, Using
Eq (\ref{eq:wstarf}) we have:

\[
W^{(*)}=SW
\]

where $S$ is a $16\times16$ matrix. This $16\times16$ matrix can
be considered as constituted of $4\times4$ blocks where each block
is again a $4\times4$ matrix and denote the blocks by $S_{ij}$.
The $S_{11}$ block is constructed by running through the points of
the first line in the first striation i.e., $(0,0),\,(0,1),\,(0,\omega)\, and\,(0,\bar{\omega})$.
So, $S_{11}$ is given by,
\begin{equation}
S_{11}=\left(\begin{array}{cccc}
Tr\left(A_{0,0}A_{0,0}^{*}\right) & Tr\left(A_{0,0}A_{0,1}^{*}\right) & Tr\left(A_{0,0}A_{0,\omega}^{*}\right) & Tr\left(A_{0,0}A_{0,\bar{\omega}}^{*}\right)\\
Tr\left(A_{0,1}A_{0,0}^{*}\right) & Tr\left(A_{0,1}A_{0,1}^{*}\right) & Tr\left(A_{0,1}A_{0,\omega}^{*}\right) & Tr\left(A_{0,1}A_{0,\bar{\omega}}^{*}\right)\\
Tr\left(A_{0,\omega}A_{0,0}^{*}\right) & Tr\left(A_{0,\omega}A_{0,1}^{*}\right) & Tr\left(A_{0,\omega}A_{0,\omega}^{*}\right) & Tr\left(A_{0,\omega}A_{0,\bar{\omega}}^{*}\right)\\
Tr\left(A_{0,\bar{\omega}}A_{0,0}^{*}\right) & Tr\left(A_{0,\bar{\omega}}A_{0,1}^{*}\right) & Tr\left(A_{0,\bar{\omega}}A_{0,\omega}^{*}\right) & Tr\left(A_{0,\bar{\omega}}A_{0,\bar{\omega}}^{*}\right)
\end{array}\right)\label{eq:S11}
\end{equation}

In this block matrix, the value of $Tr(A_{\alpha}A_{\beta}^{*})=T_{\alpha\beta}-N-2$.,
where, 
\begin{equation}
T_{\alpha\beta}=Tr[Q(\lambda_{1}^{x_{1}})Q^{*}(\lambda_{1}^{x_{1}})]+Tr[Q(\lambda_{1}^{x_{1}})Q^{*}(\lambda_{2}^{x_{2}})]+\dots+Tr[Q(\lambda_{N+1}^{x_{N+1}})Q^{*}(\lambda_{N+1}^{x_{N+1}})]\label{eq:T_ab}
\end{equation}

In $T_{\alpha\beta}$, there are $(4+1)^{2}$ trace terms, wherein
\textbf{$Q(\lambda_{i}^{x_{i}})$} and $Q^{*}(\lambda_{j}^{x_{j}})$
are from different striations, for which $Tr[Q(\lambda_{1}^{x_{1}})Q^{*}(\lambda_{2}^{x_{2}})]$
is $\frac{1}{4}$. In general there are $4(4+1)$ such trace terms
in $T_{\alpha\beta}$ . In the remaining $(4+1)$ terms, $Tr[Q(\lambda_{i}^{x_{i}})Q^{*}(\lambda_{i}^{y_{i}})]$
is either $0$ or $1$ depending on the points $\alpha$ and $\beta$.
For example,
\begin{itemize}
\item if $\alpha=(0,0)$ and $\beta=(0,0)$, $Tr[Q(\lambda_{i}^{x_{i}})Q^{*}(\lambda_{i}^{y_{i}})]=1$
for $i=1,2$ and $Tr[Q(\lambda_{i}^{x_{i}})Q^{*}(\lambda_{i}^{y_{i}})]=0$
for $i=3,4,5$. Therefore $T_{\alpha\beta}=[4(4+1)\frac{1}{4}]+1+1=7$.
So $Tr\left(A_{0,0}A_{0,0}^{*}\right)=1$. This argument holds for
all the diagonal entries.
\item if $\alpha=(0,0)$ and $\beta=(0,\omega)$, $Tr[Q(\lambda_{i}^{x_{i}})Q^{*}(\lambda_{i}^{y_{i}})]=1$
for $i=1,3$ and $Tr[Q(\lambda_{i}^{x_{i}})Q^{*}(\lambda_{i}^{y_{i}})]=0$
for $i=2,4,5$. Therefore $T_{\alpha\beta}=[4(4+1)\frac{1}{4}]+1+1=7$.
So $Tr\left(A_{0,0}A_{0,\omega}^{*}\right)=1$. Similar argument holds
for all the off-diagonal entries of the diagonal blocks. 
\end{itemize}
Therefore, all the diagonal blocks are $4\times4$ matrix whose entries
are $1$. Next, consider one block from the off-diagonal blocks- $S_{12}$
is one of them: 
\begin{equation}
S_{12}=\left(\begin{array}{cccc}
Tr\left(A_{0,0}A_{1,0}^{*}\right) & Tr\left(A_{0,0}A_{1,1}^{*}\right) & Tr\left(A_{0,0}A_{1,\omega}^{*}\right) & Tr\left(A_{0,0}A_{1,\bar{\omega}}^{*}\right)\\
Tr\left(A_{0,1}A_{1,0}^{*}\right) & Tr\left(A_{0,1}A_{1,1}^{*}\right) & Tr\left(A_{0,1}A_{1,\omega}^{*}\right) & Tr\left(A_{0,1}A_{1,\bar{\omega}}^{*}\right)\\
Tr\left(A_{0,\omega}A_{1,0}^{*}\right) & Tr\left(A_{0,\omega}A_{1,1}^{*}\right) & Tr\left(A_{0,\omega}A_{1,\omega}^{*}\right) & Tr\left(A_{0,\omega}A_{1,\bar{\omega}}^{*}\right)\\
Tr\left(A_{0,\bar{\omega}}A_{1,0}^{*}\right) & Tr\left(A_{0,\bar{\omega}}A_{1,1}^{*}\right) & Tr\left(A_{0,\bar{\omega}}A_{1,\omega}^{*}\right) & Tr\left(A_{0,\bar{\omega}}A_{1,\bar{\omega}}^{*}\right)
\end{array}\right)\label{eq:S12}
\end{equation}

Here too, the crucial step is in calculating $T_{\alpha\beta}$. In
$T_{\alpha\beta}$, there are $4(4+1)$ terms with value $\frac{1}{4}$.
Since this block is formed by the first and second lines in the first
striation, the trace term $Tr[Q(\lambda_{1}^{x_{1}})Q^{*}(\lambda_{1}^{y_{1}})]=0$.
For a given $\alpha$, with the point $\beta$ running over the points
in the second line in the first striation, $4$ times the trace terms
$Tr[Q(\lambda_{i}^{x_{i}})Q^{*}(\lambda_{i}^{y_{i}})]$ becomes $1$
for $i=2,...,5$. For $\alpha=(0,0)$ and $\beta=(1,0)$, $Tr[Q(\lambda_{i}^{x_{i}})Q^{*}(\lambda_{i}^{y_{i}})]=1$
for $i=2,4$ therefore, $Tr\left(A_{0,0}A_{1,0}^{*}\right)=1+1+(N+1)-N-2=1$.
Similarly, for $\alpha=(0,0)$ and $\beta=(1,\omega)$, $Tr[Q(\lambda_{i}^{x_{i}})Q^{*}(\lambda_{i}^{y_{i}})]=1$
for $i=3,5$ therefore $Tr\left(A_{0,0}A_{1,0}^{*}\right)=1+1+(N+1)-N-2=1$.
The remaining terms in that row are $-1$, that is $Tr\left(A_{0,0}A_{1,1}^{*}\right)=-1=Tr\left(A_{0,0}A_{1,\bar{\omega}}^{*}\right)$.
Same arguments holds for the other rows of this block, that is
the term $Tr(A_{\alpha}A_{\beta}^{*})$ becomes two times $+1$ and
the other two times $-1$. So the block matrix takes the form,
\begin{equation}
S_{12}=\left(\begin{array}{cccc}
1 & -1 & 1 & -1\\
-1 & 1 & -1 & 1\\
1 & -1 & 1 & -1\\
-1 & 1 & -1 & 1
\end{array}\right)\label{eq:S12mat}
\end{equation}

Since $Tr(A^{*}B)=Tr(AB^{*})$, if $A$ and $B$ are Hermitian matrices,
the block matrix $S_{ji}=S_{ij}{}^{T}$, where $T$ stands for transposition
. The columns of the off-diagonal blocks has equal number of $+1$
and $-1$. So, the general transformation matrix has entries which
are either $+1$ or $-1$. And rows(columns) of $S$ are orthogonal
to each other. Thus, the $S$ is a Hadamard matrix of dimension $16\times16$.
This transforms a DWF ($16\times1$ column vector) to a complex conjugate
DWF by,
\[
W^{(*)}=SW
\]

The second step, involves the calculation of $\tilde{W}$ in terms
of $W^{(*)}$, where $\tilde{W}$ is the DWF of $\tilde{\rho}$ .
We may recall, that the spin flipped density matrix $\tilde{\rho}=\left(\sigma_{y}\otimes\sigma_{y}\right)\rho^{*}\left(\sigma_{y}\otimes\sigma_{y}\right)$
as given by Eq (\ref{eq:rho-tilde}). The effect of the translation
operator $T_{\beta}=-\sigma_{y}\otimes\sigma_{y}$ on $\rho^{*}$
is to cause a shift in $W_{\alpha}^{(*)}$ by an amount $\beta$,
which depends on the basis choice for the field. From Eq (\ref{eq:translation})
the translational operator $T_{(q,p)}$in discrete phase space can
then be written as,

\begin{equation}
T_{(q,p)}=\sigma_{x}^{q_{1}}\sigma_{z}^{p_{1}}\otimes\sigma_{x}^{q_{2}}\sigma_{z}^{p_{2}}\label{eq:T-operator-two qubit}
\end{equation}

Since $\sigma_{y}=i\sigma_{z}\sigma_{x}$, we can rewrite the translational
operator $T_{\beta}=-\sigma_{y}\otimes\sigma_{y}$ by:

\begin{equation}
T_{(q,p)}=\sigma_{x}\sigma_{z}\otimes\sigma_{x}\sigma_{z}\label{eq:T-operator-Two-puali}
\end{equation}

From Eq (\ref{eq:T-operator-two qubit}) \& Eq (\ref{eq:T-operator-Two-puali})
it is clear that $q_{1}=p_{1}=q_{2}=p_{2}=1$. If we choose $(\omega,1)$
as the basis of the horizontal axis, then the basis elements for the
vertical axis also turns out to be $(\omega,1)$. Therefore $q$ and
$p$ can be expressed in terms of the basis as:

\[
q=q_{1}.e_{1}+q_{2}.e_{2}=1.\omega+1.1
\]

\begin{equation}
q=\omega+1=\bar{\omega}\label{eq:qt}
\end{equation}

and

\[
p=p_{1}.f_{1}+p_{2}.f_{2}
\]

\begin{equation}
p=\omega+1=\bar{\omega}\label{eq:pt}
\end{equation}

So, for this particular choice of bases $\beta=(\bar{\omega},\bar{\omega})$
and the translational operator $T_{\beta}=-\sigma_{y}\otimes\sigma_{y}=T_{(\bar{\omega},\bar{\omega})}$.
Therefore, 

\begin{equation}
\tilde{W}_{\alpha}=W_{\alpha+\beta}^{(*)}\label{eq:Wtilde-from-Wstar}
\end{equation}

where $\beta=(\bar{\omega},\bar{\omega})$.

\begin{table}
\protect\caption{The DWF of $W^{*}$ subjected to rigid translation effected by $\sigma_{y}\otimes\sigma_{y}$,
results in the shifting of the elements of $W^{*}$ by $\beta=(\bar{\omega},\bar{\omega})$
to yield the corresponding element of $\tilde{W}$ }
TS

$W^{*}$\qquad{}\hspace{5cm}$\tilde{W}$

\begin{tabular}{|c|c|c|c|c|}
\hline 
- - & $W_{0,\bar{\omega}}^{*}$ & $W_{1,\bar{\omega}}^{*}$ & $W_{\omega,\bar{\omega}}^{*}$ & $W_{\bar{\omega},\bar{\omega}}^{*}$\tabularnewline
\hline 
- + & $W_{0,\omega}^{*}$ & $W_{1,\omega}^{*}$ & $W_{\omega,\omega}^{*}$ & $W_{\bar{\omega},\omega}^{*}$\tabularnewline
\hline 
+ -  & $W_{0,1}^{*}$ & $W_{1,1}^{*}$ & $W_{\omega,1}^{*}$ & $W_{\bar{\omega},1}^{*}$\tabularnewline
\hline 
+ + & $W_{0,0}^{*}$ & $W_{1,0}^{*}$ & $W_{\omega,0}^{*}$ & $W_{\bar{\omega},0}^{*}$\tabularnewline
\hline 
 & HH & HV & VH & VV\tabularnewline
\hline 
\end{tabular}$\;\;\overrightarrow{\sigma_{y}\otimes\sigma_{y}}\;\;$%
\begin{tabular}{|c|c|c|c|c|}
\hline 
- - & $W_{0,\bar{\omega}}^{*}$ & $W_{\omega,0}^{*}$ & $W_{1,0}^{*}$ & $W_{0,0}^{*}$\tabularnewline
\hline 
- + & $W_{\bar{\omega},1}^{*}$ & $W_{\omega,1}^{*}$ & $W_{1,1}^{*}$ & $W_{0,1}^{*}$\tabularnewline
\hline 
+ -  & $W_{\bar{\omega},\omega}^{*}$ & $W_{\omega,\omega}^{*}$ & $W_{1,\omega}^{*}$ & $W_{0,\omega}^{*}$\tabularnewline
\hline 
+ + & $W_{\bar{\omega},\bar{\omega}}^{*}$ & $W_{\omega,\bar{\omega}}^{*}$ & $W_{1,\bar{\omega}}^{*}$ & $W_{0,\bar{\omega}}^{*}$\tabularnewline
\hline 
 & HH & HV & VH & VV\tabularnewline
\hline 
\end{tabular}
\end{table}

Therefore, from Table 1, we see that, $\tilde{W}$ can be calculated
from $W^{*}$ just by translating the elements of $W^{*}$ by $\beta=(\bar{\omega},\bar{\omega})$.
If the Wigner elements are arranged as a column vector then the this
translation is carried out by,
\begin{equation}
\tilde{W}=TW^{(*)}\label{eq:wTilde1}
\end{equation}

But the DWF of the complex conjugate state is already a Hadamard transformation
of the original DWF, that is $W^{(*)}=SW$. So this Eq can be written
as,
\begin{equation}
\tilde{W}=TSW\label{eq:wTilde2}
\end{equation}
\begin{equation}
\tilde{W}=HW\label{eq:wTilde3}
\end{equation}

It is fairly straight forward to show that the product of matrices
$T$ and $S$ is again a Hadamard matrix. We have thus illustrated
the underlying method using the one qubit and two qubit states as
examples. From these illustrations, it would be clear how the Hadamard
matrix $H$ may be pre-computed for any arbitrary multiqubit state.

\section{Conclusions}

The experimental measurement of continuous Wigner function has been
extensively reported in the quantum optics literature. In these studies,
quantum interference effects are beautifully brought out and a link
between non-classicality and negativity of the Wigner function is
transparent. However, unlike the continuous case, the discrete Wigner
function has not be as thoroughly investigated and barring some examples,
its utility is not all together clear. One of the important reasons
for this stems from the fact that the DWF is not unique and different
quantum nets give rise to different DWFs. Since the DWF depends on
the choice of the quantum net, the non-classicality or otherwise of
the reconstructed state is not obvious. Evidently, a clear interpretation
of the consequences of spin flip would require the derivation of quantities
that are independent of the choice of the underlying quantum net.
In the present work we have shown that the DWF and the spin-flipped
DWF of the multiqubit states are related through a linear transform
involving a Hadamard matrix. We have further shown that this matrix
is independent of the choice of the quantum net used in the reconstruction
of the DWF. We have illustrated this for the one and two qubit discrete
Wigner functions. Experimentally, several protocols are available
for the tomographic reconstruction of the DWF but there are no entanglement
measures defined purely in terms of discrete Wigner elements. One
way of defining entanglement measures for DWFs is use those defined
for $\rho$ and find equivalent expressions in terms of the DWF elements.
With the present results, we can readily compute bipartite concurrence
in terms of the DWF using the definition given in equation \ref{eq:Concurrence}.
With a bit of algebra, it can be shown that the matrix $R=\rho\tilde{\rho}$
may written as $R=\rho\tilde{\rho}=\underset{\alpha}{\frac{1}{4}\sum}\underset{\gamma}{\sum}W_{\alpha}\tilde{W}_{\beta}\hat{A}_{\gamma}\hat{A}_{\delta}$.
Likewise, one could also rewrite the expressions of the other tangles
in terms of the DWF. Alternately one may attempt to define altogether
SLOCC invariant measures starting from the DWF. We thus hope that
the present work may open a way to understanding entanglement directly
in terms of the DWFs. 
\begin{acknowledgments}
One of the authors(K. Srinivasan) acknowledges Indira Gandhi centre
for atomic research, DAE for the award of research fellowship. Useful
suggestions and discussions with S.Kanmani, Gururaj Kadiri and B.Radhakrishna
is hereby acknowledged by the authors. 
\end{acknowledgments}

\appendix
\bibliographystyle{apsrev}
%\bibliography{spinflip_ref}

\end{document}